\documentclass{article}
\usepackage[latin9]{inputenc}

\usepackage{amssymb,amsmath}
\usepackage[numbers]{natbib}
\usepackage{color}
\usepackage{txfonts}
\usepackage{multirow}
\usepackage{hyperref}
\usepackage{hypernat}
\usepackage{verbatim}
\usepackage{stmaryrd}
\usepackage{graphicx,float}
\usepackage[affil-it]{authblk}
\usepackage{subcaption}

\usepackage{wasysym}
\usepackage{multicol}
\usepackage{geometry}
\begin{document}
\renewcommand*{\Authfont}{\normalsize}

\title{Regular Bouncing Solutions, Energy Conditions\\ and the Brans-Dicke
Theory}

\author[1]{O. Galkina \footnote{Email: \href{mailto:olesya.galkina@cosmo-ufes.org}{olesya.galkina@cosmo-ufes.org}}}
\affil[1]{PPGFis, CCE - Universidade Federal do Esp\'\i rito Santo, zip 29075-910, Vit\'oria, ES, Brazil }

\author[2,3]{J. C. Fabris \footnote{Email: \href{mailto:julio.fabris@cosmo-ufes.org}{julio.fabris@cosmo-ufes.org}}}

\affil[2]{N\'{u}cleo Cosmo-ufes \& Departamento de F\'{\i}sica - Universidade
Federal do Esp\'\i rito Santo, zip 29075-910, Vit\'oria, ES, Brazil} \affil[3]{National
Research Nuclear University MEPhI, Kashirskoe sh. 31, Moscow 115409,
Russia}

\author[4]{F. T. Falciano \footnote{Email: \href{mailto:ftovar@cbpf.br}{ftovar@cbpf.br}}}

\author[4]{N. Pinto-Neto \footnote{Email: \href{mailto:nelsonpn@cbpf.br}{nelsonpn@cbpf.br}}}

\affil[4]{CBPF - Centro Brasileiro de Pesquisas F\'{\i}sicas, rua Xavier
Sigaud 150, zip 22290-180, Rio de Janeiro, Brazil}

\date{}

\maketitle

\begin{abstract}
In general, to avoid a singularity in cosmological models involves
the introduction of exotic kind of matter fields, for example, a scalar
field with negative energy density. In order to have a bouncing solution in classical General Relativity,
violation of the energy conditions is required. In this work, we discuss
a case of the bouncing solution in the Brans-Dicke theory with radiative
fluid that obeys the energy conditions, and with no ghosts.
\end{abstract}
\begin{multicols}{2}
\section{Introduction}

One of the main drawbacks of the standard cosmological model is the
existence of an initial singularity. Singularities are a common feature
in different applications of General Relativity (GR) when matter fields
obey reasonable energy conditions, called normal fields. Hence, the avoidance of a singularity
generally implies the introduction of exotic matter fields,
such as phantom fields (i.e., a scalar field with negative energy
density). However, there are situations where normal fields may
also lead to the avoidance of singularities if some non-trivial coupling
is introduced. This implies that the matter sector must contain more than
one component which interacts directly among themselves. Many non-singular solutions in non-minimal coupled
theories are also obtained due to the presence of
fields which appear phantom when the theory is formulated in terms of a minimally-coupled
system.

The purpose of this note is to call attention to a non-singular
model with fluids that obey the energy conditions and with no ghosts
is possible even in the most simple scalar-tensor theory, the Brans-Dicke
theory. We will essentially analyse the solutions determined by Gurevich
et al \cite{Gurevich} for a flat homogeneous and isotropic universe.
Our goal is to identify some properties of these already known solutions
which, to our knowledge, have not been studied in some of their aspects
\footnote{For a similar analysis of the solutions in the Brans-Dicke theory
see \cite{Barrow:1996kc}-\cite{Mimoso:1995ge}.}. These properties may be relevant for the construction of a coherent
and realistic cosmological model, in particular for solving the singularity
problem.

The Brans-Dicke theory of gravitation is one of the most important
alternative theories to GR, where the inverse of the gravitational
constant $G$ is replaced by a scalar filed $\phi$, which can vary
in space and time. It was developed by C. Brans and R.H. Dicke \cite{Brans-Dicke}
in order to implement the Mach's principle in a relativistic theory.
The theory has received recently much attention of the scientific
community \cite{Rossi:2019lgt}-\cite{Frion:2018oij}.

The paper is organized as follows. In section 2 we describe the system,
its equation of motion, and review the solutions for the radiative
case studied by Gurevich et al. In section 3 we analyze the bouncing
properties of the solutions. In section 4 we discuss the energy conditions,
and develop the perturbation over specific background. In section 5 we give our final remarks.

\section{The classical equations of motion and Gurevich et al Solutions}

The Brans-Dicke theory is defined by the action
\begin{eqnarray}
{\cal A}=\frac{1}{16\pi}\int d^{4}x\biggr\{\sqrt{-g}\biggr(\phi R-\frac{\omega}{\phi}\left(\nabla\phi\right)^{2}\biggl)+{\cal L}_{m}\biggl\},
\end{eqnarray}
where $\phi$ is a scalar field, ${\cal L}_{m}$ is the matter Lagrangian
and $\omega$ is a free parameter. This is the prototype of a scalar-tensor
theory where the non-minimal coupling occurs between the gravitational
term and the scalar field. The main goal of the Brans-Dicke theory
was to introduce a varying gravitational coupling through the scalar
field $\phi$. It can been seen as the first example of Galileons
and Horndesky-type theories \cite{horndesky}.

Local tests limit the value of the parameter $\omega$ to be very
large \cite{Will}, what in principle renders the theory essentially
equivalent to GR. However, extensions of the Brans-Dicke theory
leave place for a varying coupling parameter $\omega$. The Horndesky 
class of theories cover all possibilities without Ostrogradsky instabilities including 
the Brans-Dicke theory in its traditional form. 
This opens the possibility for small values of the coupling parameter in the past 
(which can be even negative), evolving to a huge value today. Also, 
the low energy effective action of string theory leads
to Brans-Dicke theory with $\omega=-1$ \cite{Duff:1994an}. Brane configurations
may allow even more negative values of $\omega$. In evoking this
connection, we have mainly in mind the domain of application of the
string effective theory which is the primordial universe.

The Brans-Dicke theory field equations read
\begin{eqnarray}
R_{\mu\nu} - \frac{1}{2}g_{\mu\nu}R & = & \frac{8\pi}{\phi}T_{\mu\nu} + \frac{\omega}{\phi^{2}}\biggr(\nabla_{\mu}\phi\nabla_{\nu}\phi-\frac{1}{2}g_{\mu\nu}\nabla_{\alpha}\phi\nabla^{\alpha}\phi\biggl)\nonumber \\
 &+& \frac{1}{\phi}(\nabla_{\mu}\nabla_{\nu}\phi-g_{\mu\nu}\Box\phi),\\
\Box\phi & = & \frac{8\pi}{3+2\omega}T,\\
\nabla_{\mu}T^{\mu\nu} & = & 0,
\end{eqnarray}
where $w$ is a constant.
For a flat FLRW metric
\begin{eqnarray}
ds^{2}=dt^{2} - a(t)^{2}(dx^{2} + dy^{2} + dz^{2}),
\end{eqnarray}
the field equations reduce to
\begin{eqnarray}
3\biggr(\frac{\dot{a}}{a}\biggl)^{2} & = & 8\pi\frac{\rho}{\phi} + \frac{\omega}{2}\biggl(\frac{\dot{\phi}}{\phi}\biggl)^{2} - 3\frac{\dot{a}}{a}\frac{\dot{\phi}}{\phi},\\
2\frac{\ddot{a}}{a} + \biggr(\frac{\dot{a}}{a}\biggl)^{2} & = & - 8\pi\frac{p}{\phi} - \frac{\omega}{2}\biggl(\frac{\dot{\phi}}{\phi}\biggl)^{2} - \frac{\ddot{\phi}}{\phi} - 2\frac{\dot{a}}{a}\frac{\dot{\phi}}{\phi},\\
\ddot{\phi} + 3\frac{\dot{a}}{a}\dot{\phi} & = & \frac{8\pi}{3+2\omega}(\rho-3p),\\
\dot{\rho} + 3\frac{\dot{a}}{a}(\rho+p) & = & 0.
\end{eqnarray}

Gurevich et al \cite{Gurevich} determined the general solution for the cosmological
isotropic and homogeneous flat universe with a perfect fluid with
an equation of state $p=\alpha\rho$, where $\alpha$ is a constant such
that $0\leq\alpha\leq1$. The general solution for $\omega>-\frac{3}{2}$
(a case where the energy conditions for the scalar field are satisfied)
reads
\begin{eqnarray}
a(\theta) & = & a_{0}(\theta-\theta_{+})^{r_{+}}(\theta-\theta_{-})^{r_{-}},\\
\phi(\theta) & = & \phi_{0}(\theta-\theta_{+})^{s_{+}}(\theta-\theta_{-})^{s_{-}},
\end{eqnarray}
with the definitions,
\begin{eqnarray}
r_{+} & = & \frac{\omega}{3\biggl[\sigma\mp\sqrt{1+\frac{2}{3}\omega}\biggr]},\quad r_{-}=\frac{\omega}{3\biggl[\sigma\pm\sqrt{1+\frac{2}{3}\omega}\biggr]},\\
s_{+} & = & \frac{1\mp\sqrt{1+\frac{2}{3}\omega}}{\sigma\mp\sqrt{1+\frac{2}{3}\omega}},\quad s_{-}=\frac{1\pm\sqrt{1+\frac{2}{3}\omega}}{\sigma\pm\sqrt{1+\frac{2}{3}\omega}},
\end{eqnarray}
where $\sigma=1+\omega(1-\alpha)$, and $a_0 ,\phi_0 ,\theta_{\pm}$ are arbitrary constants, with $\theta_+ >\theta_-$. The time coordinate $\theta$
is connected with the cosmic time $t$ by
\begin{eqnarray}
dt=a^{3\alpha}d\theta.
\end{eqnarray}

For $\omega<-\frac{3}{2}$, where there is violation of
the energy conditions for the scalar field in the Einstein frame, as it will be discussed
below, the solutions read
\begin{align}
a & =a_{0}\Big[\left(\theta+\theta_{-}\right)^{2}+\theta_{+}^{2}\Big]^{\big(1+(1-\alpha)\omega\big)/{A}}e^{ \pm 2 f(\theta)/A}\ ,\label{eq:general solution}\\
\phi & =\phi_{0}\left[\left(\theta+\theta_{-}\right)^{2}+\theta_{+}^{2}\right]^{{(1-3\alpha)}/{A}}e^{ \pm 6\left(1-\alpha\right) f(\theta)/A}\ , \label{eq:general solution phi}
\end{align}
where
\begin{align}
f(\theta)&= \sqrt{\frac{2|\omega|-3}{3}}\arctan\left(\frac{\theta+\theta_{-}}{\theta_{+}}\right)\label{eq:general solution f}\\
A&=2\left(2-3\alpha\right)+3\omega\left(1-\alpha\right)^2\label{eq:general solution A}
\end{align}

In the case $\omega>-\frac{3}{2}$, the condition to have a regular
bounce can be expressed by requiring $r_{+}<0$ (the scale factor
is infinite at one asymptote), $r_{+}+r_{-}>0$ (the scale factor
is infinite at another asymptote) and $3\alpha r_{+}+1<0$ (the cosmic time
varies from $-\infty$ to $+\infty$). These conditions imply that
a regular bounce may be obtained for $\frac{1}{4}<\alpha<1$ and
$-\frac{3}{2}<\omega\leq-\frac{4}{3}$. The case $\alpha=1$ is quite
peculiar, and contains no bounce \cite{Brando:2018kic}.

We will be interested here mainly in a scenario for the early universe.
Thus, we will consider in detail the radiative universe. The Gurevich
et al solution for the radiative case ($p=\frac{1}{3}\rho$) is given
by the following expressions
\begin{itemize}
\item $\omega>-\frac{3}{2}$:
\begin{eqnarray}
a(\eta) & = & a_{0}(\eta-\eta_{+})^{\frac{1}{2}(1\pm r)}(\eta-\eta_{-})^{\frac{1}{2}(1\mp r)},\label{b1}\\
\phi(\eta) & = & \phi_{0}(\eta-\eta_{+})^{\mp r}(\eta-\eta_{-})^{\pm r};\label{b2}
\end{eqnarray}
\item $\omega<-\frac{3}{2}$:
\begin{eqnarray}
a(\eta) & = & a_{0}[(\eta+\eta_{-})^{2}+\eta_{+}^{2}]^{\frac{1}{2}}e^{\pm\frac{1}{\sqrt{\frac{2}{3}|\omega|-1}}\arctan\frac{\eta+\eta_{-}}{\eta_{+}}},\label{eq:scale factor negative}\\
\phi(\eta) & = & \phi_{0}e^{\mp\frac{2}{\sqrt{\frac{2}{3}|\omega|-1}}\arctan\frac{\eta+\eta_{-}}{\eta_{+}}}.\label{eq:scalar field negative}
\end{eqnarray}
\end{itemize}
In these expressions,
\begin{eqnarray}
r=\frac{1}{\sqrt{1+\frac{2}{3}\omega}},
\end{eqnarray}
$\eta$ is the conformal time and $\eta_{\pm}$ are constants such
that $\eta_{+}>\eta_{-}$.

If we perform a conformal transformation of the Brans-Dicke action
such that $g_{\mu\nu}=\phi^{-1}\tilde{g}_{\mu\nu}$, we re-express
it in the so-called Einstein's frame
\begin{eqnarray}
{\cal A}=\frac{1}{16\pi}\int d^{4}x\biggr\{\sqrt{-\tilde{g}}\biggr[\tilde{R}-\biggr(\omega+\frac{3}{2}\biggl)\frac{\left(\nabla\phi\right)^{2}}{\phi^{2}}\biggl]+{\cal L}_{m}\biggl\}.
\end{eqnarray}
Thus, in the Einstein frame, $\omega>-\frac{3}{2}$ corresponds to
an ordinary scalar field with positive energy density, while for $\omega<-\frac{3}{2}$,
the kinetic term of the scalar field changes sign, and it becomes
a phantom field with negative energy density. Remember
that the radiative fluid is conformal invariant.

\section{Analysis of the Solutions}

For $\omega\geq0$ the scale factor displays an initial singularity
followed by expansion, reaching $a\rightarrow\infty$ as $\eta\rightarrow\infty$.
Note that the radiative universe of GR characterised
by
\begin{eqnarray}
a\propto\eta,
\end{eqnarray}
can be recovered from the above solutions if
$\eta_{\pm}=0$, in the limit $\omega\rightarrow\infty$ when $\eta_{+}=\eta_{-}$,
or in the asymptotic limit $\eta\rightarrow\infty$. 

The GR behaviour of the scale factor is also achieved for $\omega=0$.
However, in this case, the scalar field (the inverse of the gravitational
coupling) varies with time, and its variation depends essentially on
the sign in the exponent in Eqs. (\ref{b1})(\ref{b2}). For the upper
sign, we find
\begin{eqnarray}
a(\eta) & = & a_{0}(\eta-\eta_{+}),\\
\phi(\eta) & = & \phi_{0}\frac{\eta-\eta_{-}}{\eta-\eta_{+}},
\end{eqnarray}
and the scalar field decreases monotonically from infinite to a constant
(positive) value as the universe evolves. For the lower sign, the
behaviour of the functions are given by
\begin{eqnarray}
a(\eta) & = & a_{0}(\eta-\eta_{-}),\\
\phi(\eta) & = & \phi_{0}\frac{\eta-\eta_{+}}{\eta-\eta_{-}},
\end{eqnarray}
and the scalar field increases monotonically from an infinite negative
value to a constant positive value as the universe evolves: initially
there is a repulsive gravitational phase. This can be considered as a Big Rip type singularity since it occurs when $a\rightarrow\infty$ at finite proper time.

Bounce solutions can be obtained from the Gurevich et al solutions in
the radiative case if the lower sign is
chosen in Eqs. (\ref{b1})(\ref{b2}) for $-\frac{3}{2}<\omega<0$.
However, there is a singularity at $\eta=\eta_{+}$ for $-\frac{4}{3}<\omega<0$ at $\eta=\eta_{+}$,
even if the scale factor diverges at this point. On the other hand, if $-\frac{3}{2}<\omega\leq-\frac{4}{3}$,
the bounce solutions are always regular, with no curvature singularity\footnote{Note that the gravitational coupling diverges, but only at infinite cosmic time, where the scale factor is also infinite. One can expect that instabilities (due to the anisotropic perturbations) do not develop since, in this situation, anisotropies are suppressed as they decay fast when the scale factor increases. This kind of instabilities may be very relevant, however, if there is a change of sign in the gravitational coupling at finite scale factor, as in the case of Ref. \cite{Starobinsky}.}.
In this last case, there are two possible scenarios (thanks to time
reversal invariance):
\begin{enumerate}
\item A universe that begins at $\eta=\eta_{+}$ with $a\rightarrow\infty$,
with an infinite value for the gravitational coupling ($\phi=0$),
evolving to the other asymptotic limit with $a\rightarrow\infty$
but with $\phi$ constant and finite;
\item The reversal behaviour occurs for $-\infty<\eta<-\eta_{+}$.
\end{enumerate}
In both cases, the cosmic times ranges $-\infty<t<\infty$. The dual solution in the Einstein frame for $-\frac{3}{2}<\omega\leq-\frac{4}{3}$  is given by $b(\eta)=b_{0}(\eta-\eta_{+})^{1/2}(\eta-\eta_{-})^{1/2}$ (with $b=\phi^{1/2}a$) and contains an initial singularity. This can be considered as a specific case of "conformal continuation" in the scalar-tensor gravity proposed in \cite{Bronnikov:2002kf}.

For the special case $\omega=-\frac{4}{3}$ there is still no singularity
if we choose the lower sign. In this case, the scale factor and the
scalar field behaves
\begin{eqnarray}
a(\eta) \propto  \frac{(\eta-\eta_{-})^{2}}{\eta-\eta_{+}},\quad
\phi(\eta) \propto  \biggr(\frac{\eta-\eta_{+}}{\eta-\eta_{-}}\biggl)^{3}.
\end{eqnarray}
If $-\infty<\eta<\eta_{+}$ the universe begins with $a\rightarrow\infty$,
with $\phi$ constant and finite, while in the remote future $a\rightarrow\infty$
and $\phi=0$. If we choose the interval $\eta_{+}\leq\eta<\infty$,
the scenario is reversed, and we get the possibility to have a constant gravitational
coupling today.

For $\omega=-\frac{4}{3}$ and the upper sign the solutions exhibit
an initial singularity:
\begin{eqnarray}
a(\eta) & \propto & \frac{(\eta-\eta_{+})^{2}}{\eta-\eta_{-}},\\
\phi(\eta) & \propto & \biggr(\frac{\eta-\eta_{-}}{\eta-\eta_{+}}\biggl)^{3}.
\end{eqnarray}

Similar features for the scale factor and the scalar field are reproduced
for $\omega<-\frac{3}{2}$. However the scalar field has a phantom
behaviour as already stated.

\section{Energy Conditions and Perturbations}

An important aspect of these solutions concerns the energy conditions.
In general in order to have a bouncing solution, violation of the
energy conditions is required. The strong and null energy conditions
in General Relativity are given by
\begin{eqnarray}
\frac{\ddot{a}}{a} & = & -\frac{4\pi G}{3}(\rho+3p)>0,\label{ec1}\\
-2\frac{\ddot{a}}{a}+2\biggr(\frac{\dot{a}}{a}\biggl) & = & 8\pi G(\rho+p)>0.\label{ec2}
\end{eqnarray}

\begin{figure}[!t]
\begin{minipage}[t]{0.24\linewidth}%
\includegraphics[width=1\linewidth]{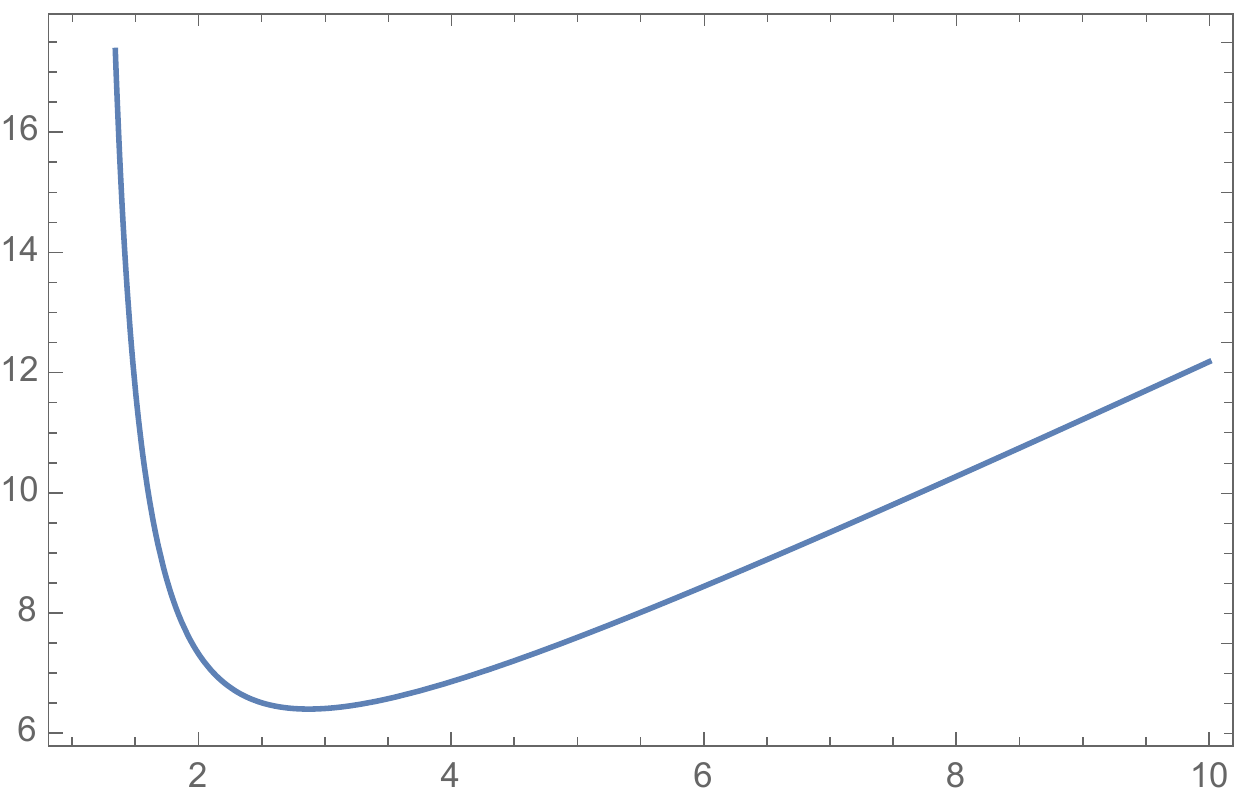} %
\end{minipage}\hfill{}%
\begin{minipage}[t]{0.24\linewidth}%
\includegraphics[width=1\linewidth]{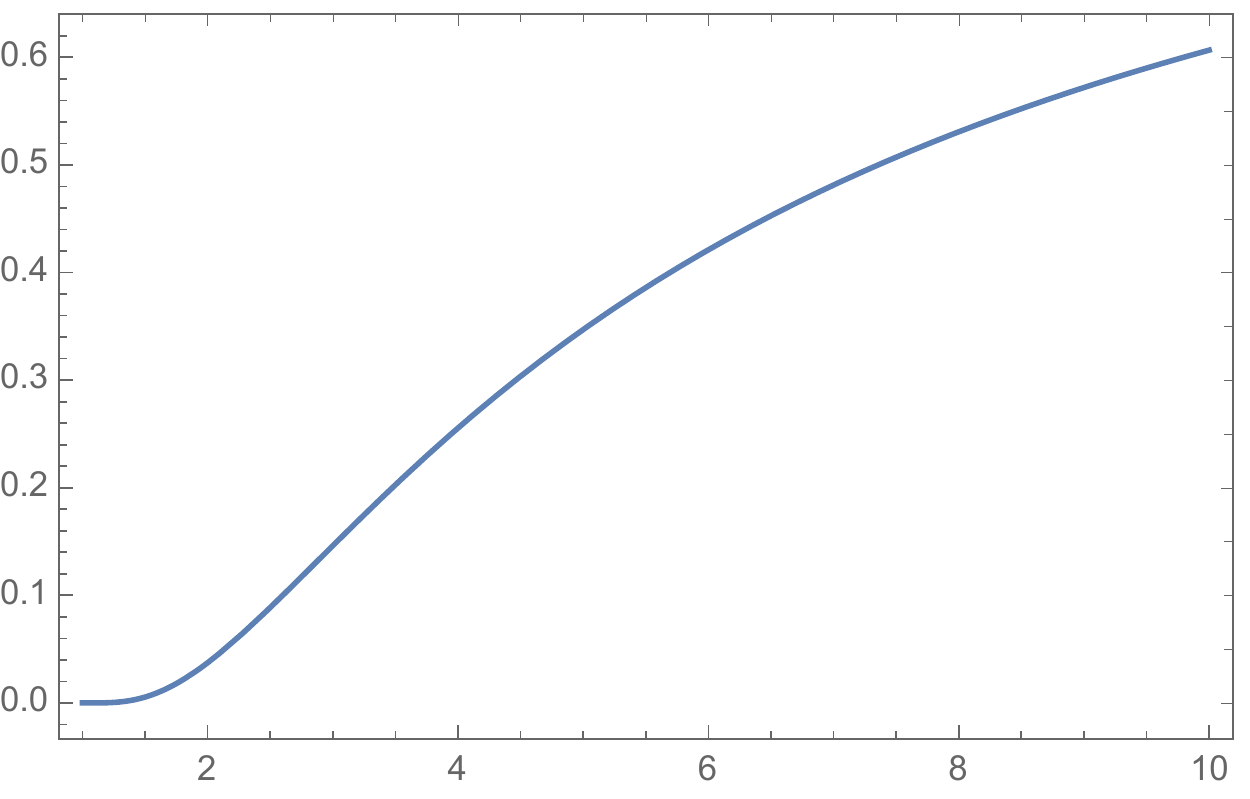} %
\end{minipage}\hfill{}%
\begin{minipage}[t]{0.24\linewidth}%
\includegraphics[width=1\linewidth]{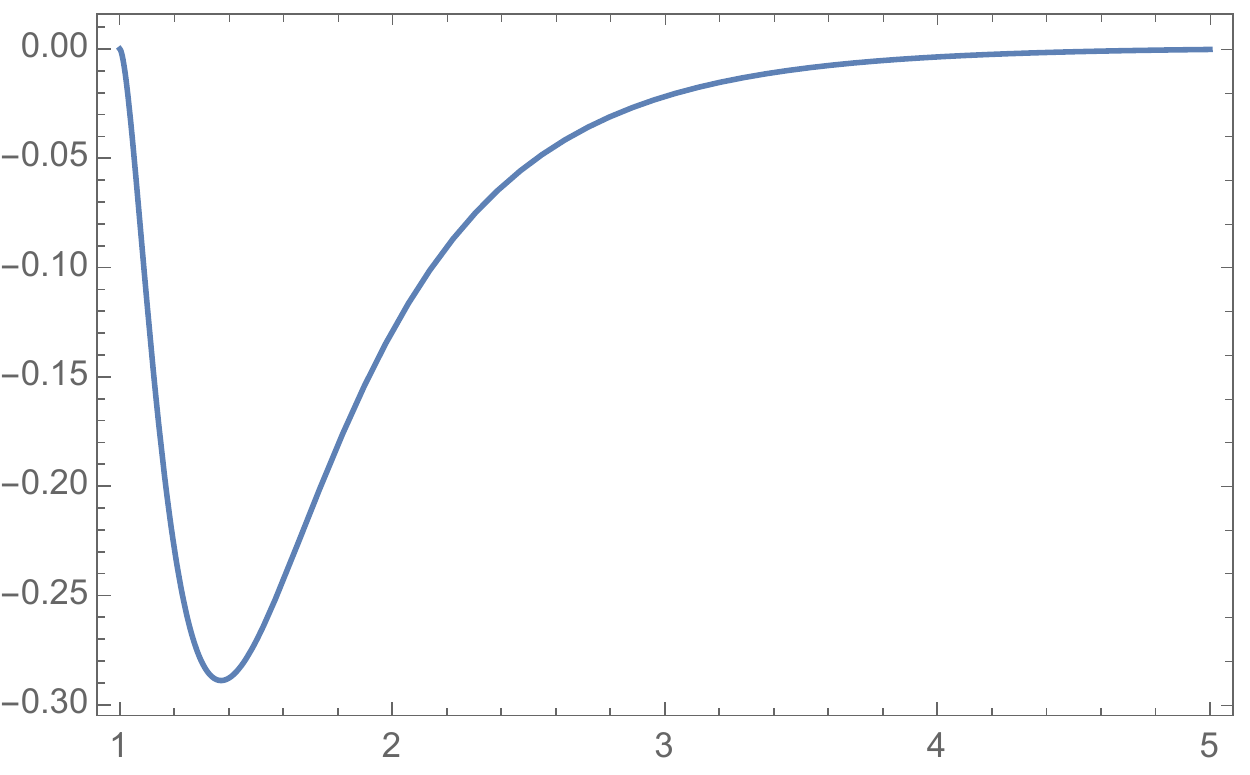} %
\end{minipage}\hfill{}%
\begin{minipage}[t]{0.24\linewidth}%
\includegraphics[width=1\linewidth]{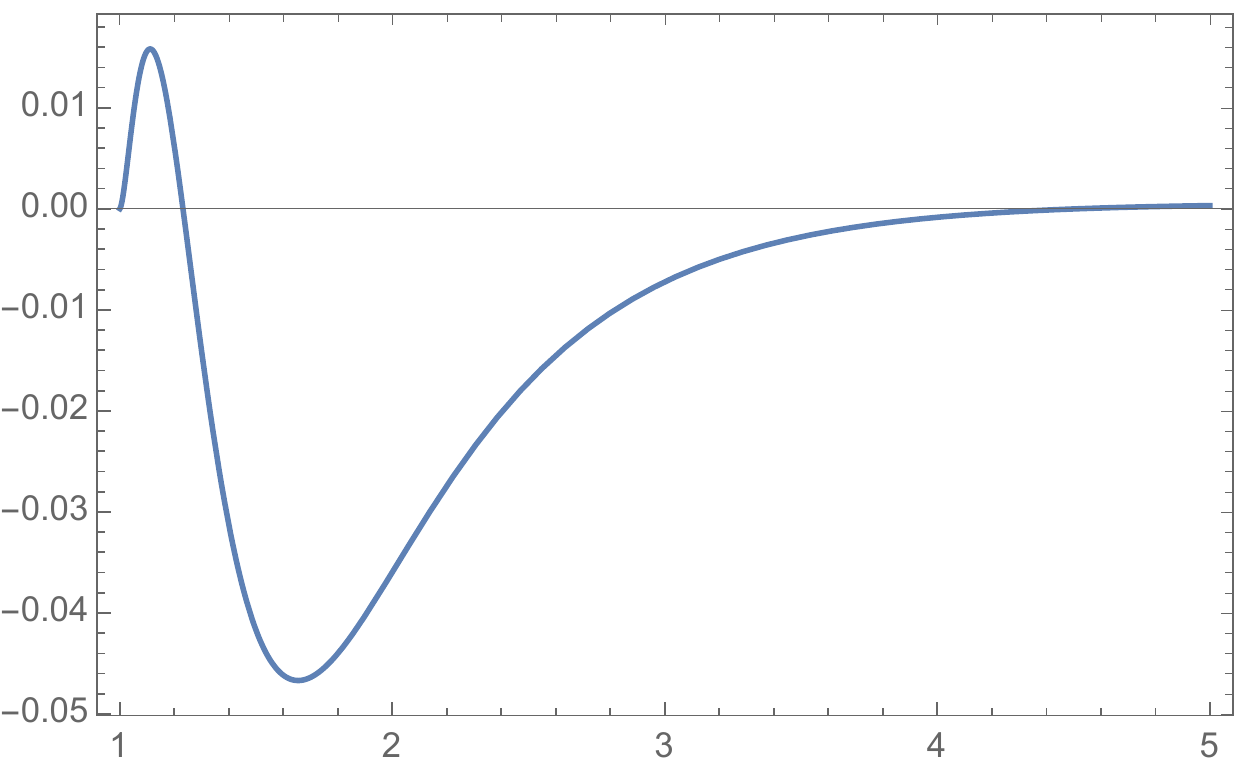} %
\end{minipage}\hfill{}\caption{\label{fig:energy conditions}Behaviour of the scale factor (left),
the scalar field (center left), of the \char`\"{}effective\char`\"{}
strong energy condition (center right) and \char`\"{}effective\char`\"{}
null energy condition (right) for $\omega=-1.43$ lower sign.}
\end{figure}

In order to use the energy condition in this form the Brans-Dicke
theory must be reformulated in the Einstein frame. It is easy to verify
that both energy conditions are satisfied as far as $\omega<-\frac{3}{2}$.
This is consistent with the fact that in the Einstein frame the cosmological
scenarios are singular unless $\omega<-\frac{3}{2}$. On the other
hand, in the original Jordan frame there are non singular models if
$-\frac{3}{2}<\omega<-\frac{4}{3}$. But in this range the scalar
field obeys the energy condition. The effects leading to the avoidance
of the singularity come from the non-minimal coupling. We plot the
\char`\"{}effective\char`\"{} energy condition, represented in the
left-hand side of Eqs. (\ref{ec1})(\ref{ec2}), taking into account
the effects of the non-minimal coupling. If we consider only the left-hand
side of the relations Eqs. (\ref{ec1})(\ref{ec2}), the effects of
the interaction due to the non-minimal coupling are included, and
the energy conditions can be violated even if the matter terms do
not violated them. In Fig. \ref{fig:energy conditions} we show the
expressions for these relations for some values of $\omega$.

It is interesting to notice that, for the most usual fluids employed in cosmology, the case of the radiative fluid is
the only one where the possibility of obtaining a singularity-free
scenario preserving the energy conditions is possible, at least in
the Brans-Dicke theory \footnote{Also with a flat spatial section. For a non-flat universe, a singularity-free
scenario can be obtained even in General Relativity if the strong
energy condition (but not necessarily the null energy condition) is
violated.}. For a matter fluid ($p=0$), the scale factor can be expressed in
terms of the cosmic time and behaves, according the Gurevich et al
solution, as
\begin{eqnarray}
a(t)=a_{0}(t-t_{+})^{r_{\pm}}(t-t_{-})^{r_{\mp}},\quad r_{\pm}=\frac{1+\omega\pm\sqrt{1+\frac{2}{3}\omega}}{4+3\omega},
\end{eqnarray}
$t_{\pm}$ being integration constants such that $t_{+}>t_{-}$. There
is a singular bounce for negative values of $\omega$ . In their work,
Gurevich et al does not display explicitly the solution for a vacuum
equation of state ($p=-\rho$) but it can be deduced from a general
expression they write down. For $p=-\rho$ the general solution reduces
to
\begin{eqnarray}
a(\theta) & = & a_{0}(\theta-\theta_{+})^{s_{\pm}}(\theta-\theta_{-})^{s_{\mp}},\quad
\end{eqnarray}
\begin{eqnarray}
s_{\pm}=\frac{1+2\omega\pm\sqrt{1+\frac{2}{3}\omega}}{2(5+6\omega)},
\end{eqnarray}
where $\theta$ is a parametric time, which is connected to cosmic
time through the relation $dt=a^{-3}d\theta$. As in the pressureless
matter case, bounce solutions exist for negative $\omega$, but they
are singular. Of course, in both pressureless and cosmological constant
cases singularity free solutions are possible if $\omega<-\frac{3}{2}$
but this implies a phantom scalar field.

Now, let us turn to perturbations. Using the synchronous coordinate
condition and particularising the expressions for a radiative fluid,
the perturbed equations read
\begin{eqnarray}
\ddot{h}+2H\dot{h}& = & \frac{16\pi}{\phi}(\delta-\lambda)+2\ddot{\lambda}+4\frac{\dot{\phi}}{\phi}(1+\omega)\dot{\lambda},
\end{eqnarray}
\begin{eqnarray}
\ddot{\lambda}+\biggr(3H+2\frac{\dot{\phi}}{\phi}\biggl)\dot{\lambda}+\frac{k^{2}}{a^{2}}\lambda & = & \frac{\dot{\phi}}{\phi}\frac{\dot{h}}{2},\\
\dot{\delta}+\frac{4}{3}\biggr(\theta-\frac{\dot{h}}{2}\biggl) & = & 0,\\
\dot{\theta}+H\theta & = & \frac{k^{2}}{4a^{2}}\delta.
\end{eqnarray}
In these expressions we have
\begin{eqnarray}
h=\frac{k_{kk}}{a^{2}},\quad\delta=\frac{\delta\rho}{\rho},\quad\lambda=\frac{\delta\phi}{\phi},\quad\theta=\partial_{i}\delta u^{i}.
\end{eqnarray}
Moreover, $k$ is the wavenumber coming from the Fourier decomposition
and $H$ is the Hubble function.

The evolution of scalar perturbations in the Brans-Dicke theory has
been studied in Ref. \cite{Baptista:1996rr}, and some features connected
with the Gurevich et al solutions have been displayed in Ref. \cite{Baptista:1989dv}.
For the bouncing regular solutions analysed here, it is natural to
implement the Bunch-Davies vacuum state as the initial condition.
However, it is known that in bounce scenario a flat or almost flat
spectrum requires a matter dominant period in the contraction phase.
This is not obviously the case for the regular Gurevich et al solutions
which is verified for a radiative fluid.

\begin{figure}[!t]
\begin{minipage}[t]{0.3\linewidth}%
\includegraphics[width=1\linewidth]{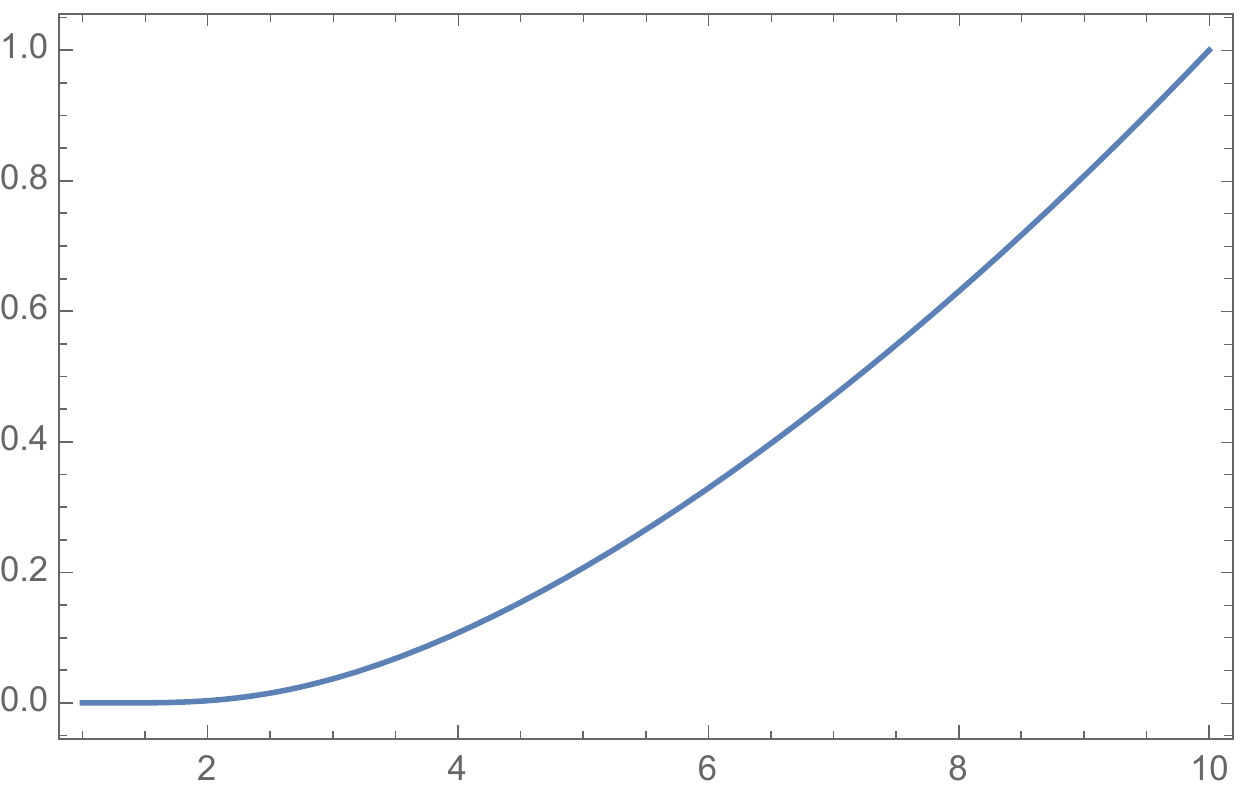} %
\end{minipage}\hfill{}%
\begin{minipage}[t]{0.3\linewidth}%
\includegraphics[width=1\linewidth]{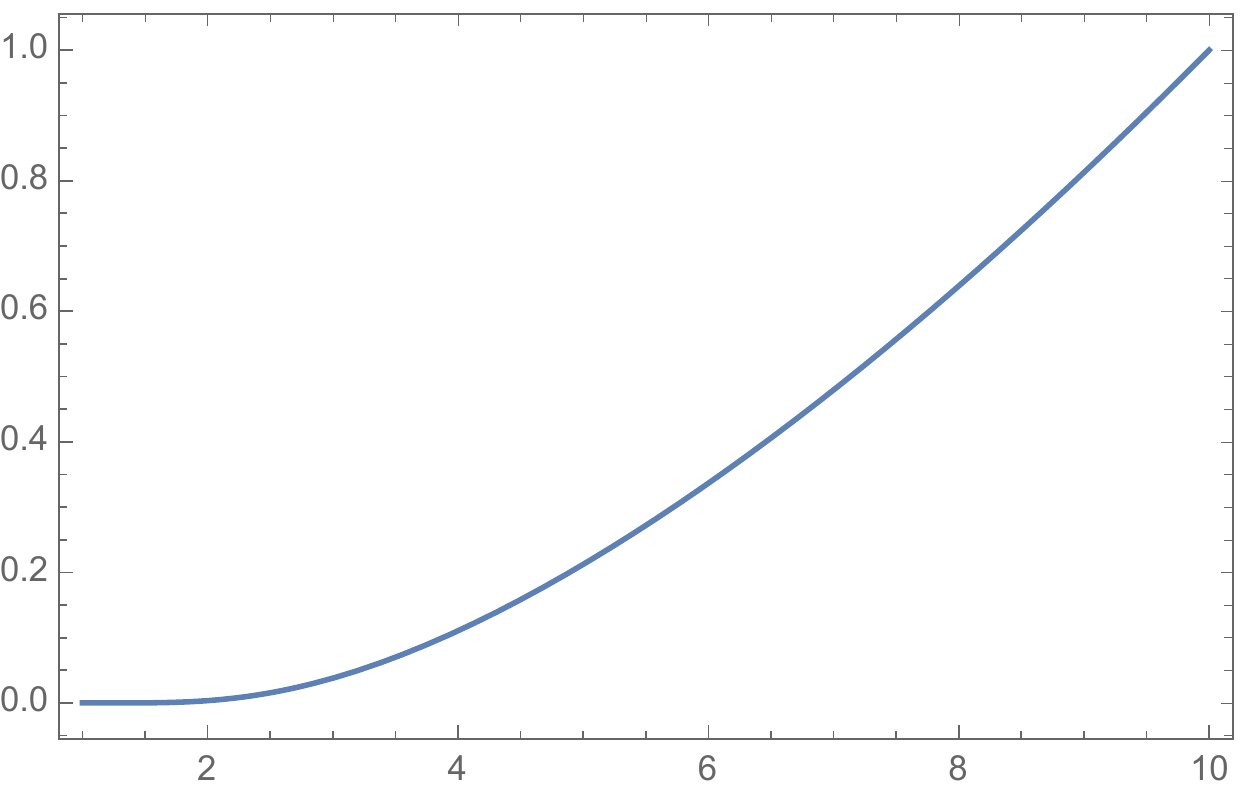} %
\end{minipage}\hfill{}%
\begin{minipage}[t]{0.33\linewidth}%
\includegraphics[width=0.91\linewidth]{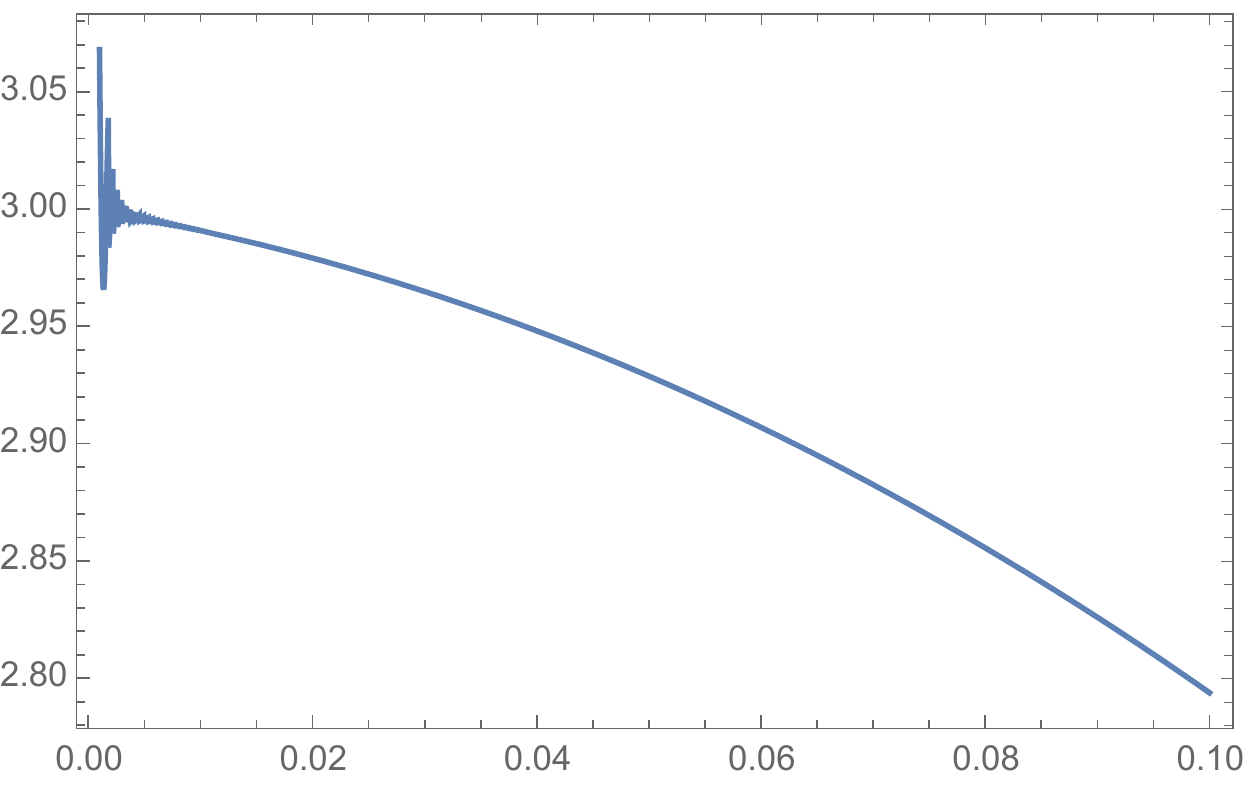} %
\end{minipage}\hfill{}\caption{\label{fig:pertu}The behaviour of the density contrast is displayed
for $k=0.01$ (left panel) and for $k=0.1$ (center panel). The normalization has been chosen such that the final density contrast is equal to one. In the right
panel is shown the dependence of the spectral index $n_{s}$ on the
wavenumber $k$. All the figures were obtained for $\omega=-1.43$
lower sign.}
\end{figure}

In Fig. \ref{fig:pertu} we display the evolution for the density
contrast for $k=0.01$ and $k=0.1$ (in the units of the current Hubble
scale), as well as the dependence of the spectral index $n_{s}$ as
a function of the wavenumber $k$. The spectral index is defined as
usual
\begin{eqnarray}
\Delta=k^{3}\delta_{k}^{2}=k^{n_{s}-1}.
\end{eqnarray}
We display the evolution of the perturbations and the dimensionless
power spectrum which exhibits a clear disagreement with the observations
(compare with similar results obtained in Ref. \cite{Vitenti:2011yc}).
Since the model studied here requires a single radiative fluid such
somehow negative result could be expected from the beginning.

\section{Discussion}

In this paper we have shown that 
regular bounce solutions without any phantom field, even in the Einstein frame, can arise in Brans-Dicke theories containing
fluids obeying the equation of state $p=\alpha \rho$ if 
$\frac{1}{4}\leq\alpha<1$, and a Brans-Dicke parameter $\omega$ lying in the interval
$-\frac{3}{2}\leq\omega\leq-\frac{4}{3}$, enlarging the parameter space in which such cosmological models
can emerge in this class of theories. 

We analysed in detail the radiative case with $\alpha=\frac{1}{3}$.
A bounce can be obtained if we choose the lower sign in Eqs. (\ref{b1})(\ref{b2})
for $-\frac{3}{2}<\omega<0.$ Moreover, for $-\frac{3}{2}<\omega\leq\frac{4}{3}$
the bounce is regular with no curvature singularity, but for $-\frac{4}{3}<\omega<0$
there is a singularity at $\eta=\eta_{+}$, even if the scale factor diverges
at this point. In the case of $\omega=-\frac{4}{3}$ there is still
no singularity if we choose the lower sign, and there is an initial
singularity for the upper sign. The solutions 
Eqs. (\ref{eq:scale factor negative})(\ref{eq:scalar field negative})
with $\omega<-\frac{3}{2}$ have a similar behaviour, but with a phantom
field in the Einstein frame.

It is generally expected that the violation of the energy conditions is required in
order to have classical bounce solutions, even in the non-minimal coupling case: in this situation,
phantom fields would appear in the Einstein frame. We discussed this point in detail for the case of
the radiative fluid in the Brans-Dicke theory (with a flat spatial
section), where we have shown that it is possible to obtain non-singular solutions preserving
the energy conditions even in the Einstein frame, and we have shown that this property holds
for any Brans-Dicke theory in which $\frac{1}{4}\leq\alpha<1$, and 
$-\frac{3}{2}\leq\omega\leq-\frac{4}{3}$. This generalization allows the possibility of constructing
more involved and realistic regular bouncing solutions, in which the power spectrum of cosmological perturbations
could be in accordance with present observations. This is one of our goals of our future investigations in this subject. 

\end{multicols}

\begin{multicols}{2}

\section*{Acknowledgments}

The authors would like to thank and acknowledge financial support
from the Conselho Nacional de Desenvolvimento Cient\'{\i}fico e Tecnol\'ogico (CNPq,
Brazil), and the Funda\c{c}\~ao de Apoio \`a Pesquisa e Inova\c{c}\~ao do Esp\'{\i}rito Santo (FAPES, Brazil). This study was financed in part by the Coordena\c c\~ao de Aperfei\c{c}oamento de Pessoal de N\'\i vel Superior - Brasil (CAPES) - Finance Code 001.

\end{multicols}
\end{document}